\documentclass{PoS}                              

\usepackage[latin1]{inputenc}                    
\usepackage{amsmath}                             

\graphicspath{{./}{Figures/}}                    
\newcommand{\imag}{\mbox{i}}                     
\newcommand{\mps}{m_{\pi}}                       
\newcommand{\dlangle}{\langle\!\langle}          
\newcommand{\drangle}{\rangle\!\rangle}          

\title{Nucleon form factors from high statistics mixed-action
  calculations with 2+1 flavors}

\ShortTitle{Nucleon form factors}

\author{LHPC Collaboration: \speaker{W.~Schroers}$^{ab}$,
  J.D.~Bratt$^c$, R.G.~Edwards$^d$, M.~Engelhardt$^e$,
  G.T.~Fleming$^f$, Ph.~H{\"a}gler$^g$, H.W.~Lin$^d$, M.F.~Lin$^c$,
  H.B.~Meyer$^c$, B.~Musch$^g$, J.W.~Negele$^c$, K.~Orginos$^h$,
  A.V.~Pochinsky$^c$, M.~Procura$^c$, D.B.~Renner$^i$,
  D.G.~Richards$^d$, S.N.~Syritsyn$^c$, and A.P.~Walker-Loud$^h$ \\
  \llap{$^a$} Institute of Physics, Academia Sinica, Taipei 115,
  Taiwan
  (R.O.C.) \\
  \llap{$^b$} Department of Physics, Center for Theoretical Sciences,
  National Taiwan University, Taipei 10617, Taiwan (R.O.C.) \\
  \llap{$^c$} Center for Theoretical Physics, Massachusetts Institute
  of
  Technology, Cambridge, MA 02139 \\
  \llap{$^d$} Thomas Jefferson National Accelerator Facility, Newport
  News, VA 23606 \\
  \llap{$^e$} Department of Physics, New Mexico State University, Las
  Cruces, NM 88003-0001 \\
  \llap{$^f$} Sloane Physics Laboratory, Yale University, New Haven,
  CT
  06520 \\
  \llap{$^g$} Institut f\"ur Theoretische Physik T39,
  Physik-Department der TU M\"unchen, James-Franck-Stra\ss{}e, D-85747
  Garching, Germany
  \\
  \llap{$^h$} Department of Physics, College of William and
  Mary,P.O. Box 8795, Williamsburg VA 23187-8795 \\
  \llap{$^i$} NIC/DESY, D-15738 Zeuthen, Germany \\ }

\abstract{We present new high-statistics results for nucleon form
  factors at pion masses of approximately 290, 350, 500, and 600 MeV
  using a mixed action of domain wall valence quarks on an improved
  staggered sea. We perform chiral fits to both vector and axial form
  factors and compare our results to experiment.}

\FullConference{The XXVII International Symposium on Lattice Field
  Theory - LAT2009\\
  July 26-31 2009\\
  Peking University, Beijing, China}

\begin{document}

\section{\label{sec:introduction}Introduction}
The proton and the neutron are the fundamental building blocks of our
world. They form heavier nuclei and thus the basis for atoms and are
the only known source of stable baryonic matter. Their structure can
be studied by scattering leptons off nuclei and the most basic
observables obtained from these processes are the nucleon form
factors. The electromagnetic form factors are Lorentz-scalars which
parametrize the matrix element of the electromagnetic current between
two nucleon states at different momentum:
\begin{equation}
  \label{eq:nucl-ff-def}
  \langle p'\vert \bar{q}\gamma^\mu q \vert p\rangle =
  \dlangle\gamma^{\mu}\drangle F_1(Q^2) + \frac{\imag}{2
    m_N}\dlangle\sigma^{\mu\alpha}\drangle \Delta_\alpha F_2(Q^2)\,,
\end{equation}
where $\dlangle {\cal X}\drangle\equiv \bar{u}(p'){\cal X}u(p)$ and
$Q^2\equiv -\Delta^2=-(p'-p)^2$. $m_N$ refers to the nucleon mass, and
we only consider the isovector combination, i.e.~the difference
between u- and d-quark currents. In this case, contributions from
disconnected diagrams cancel due to isospin symmetry. A different
parametrization called the Sachs form factors, $G_E(Q^2)$ and
$G_M(Q^2)$, is often used in the literature. We will discuss those in
our upcoming paper~\cite{LHPC:2009pr}.

In a similar way, the axial current can be parametrized in terms of
two form factors. For the isovector form factors, i.e.~the proton
minus the neutron combination, they are called the axial form factor,
$G_A(Q^2)$, and the pseudoscalar form factor, $G_P(Q^2)$:
\begin{equation}
  \label{eq:ax-ff-def}
  \langle p'\vert \bar{q}\gamma_5\gamma^\mu q\vert p\rangle = \dlangle
  \gamma^{\mu}\gamma_5\drangle G_A{Q^2} + \frac{1}{2
    m_N}\Delta^\mu\dlangle \gamma_5\drangle G_P(Q^2)\,.
\end{equation}
The current work is based on mixed action calculations using two
flavors of dynamical asqtad sea quarks~\cite{Allton:2008pn} and domain
wall valence quarks. In previous years, we have reported on several
other nucleon structure observables using this technology, see
e.g.~\cite{Hagler:2007xi, WalkerLoud:2008bp, Schroers:2005rm,
  Syritsyn:2009np}. We have studied form factors on full DWF lattices
in Ref.~\cite{Syritsyn:2009mx} which also includes comparison to the
work reported here. For a concise review of key results, we refer to
Ref.~\cite{Meifeng:2009lat}. For recent results from other groups, see
Ref.~\cite{Lin:2008uz}, and for recent reviews of the field
Ref.~\cite{Zanotti:2008zm, Renner:2009lat}.

The current report focuses on selected results for nucleon form
factors and uses several technological updates. The final report using
these improvements will be published soon~\cite{LHPC:2009pr} and will
include several other major observables such as moments of generalized
parton distributions and structure functions.

\section{\label{sec:lattice-technology}Lattice technology}
As in our previous studies we employ the asqtad action for the sea
quarks and the domain wall (DWF) action for the valence quarks. In
addition, we also add one lighter mass to our data set. The tuning of
the quarks masses and the choice of parameters have been discussed in
Ref.~\cite{Hagler:2007xi}. The lattice spacing for all data sets
corresponds to $a=0.124$ fm with an uncertainty of $2\%$, see
Ref.~\cite{Aubin:2004wf}. This yields a physical volume $V=(2.5$
fm$)^3$ on the $20^3$ and $V=(3.5$ fm$)^3$ on the $28^3$ lattices. The
resulting physical values of the nucleon and pion masses are needed
for our calculation and have been determined previously in
Ref.~\cite{WalkerLoud:2008bp}. Table~\ref{tab:phys-pars} lists these
numbers.
\begin{table}[htb] \centering
  \begin{tabular}[c]{*{5}{c|}c}
    \hline\hline
    \textbf{Light} \strut $\mathbf{m_{\mbox{\tiny sea}}^{\mbox{\tiny
          Asqtad}}}$ & \textbf{Volume} & $\mathbf{(am)_\pi}$ &
    $\mathbf{(am)_N}$ & $\mathbf{\mps}$ \textbf{/ MeV} &
    $\mathbf{m_N}$ \textbf{/ MeV} \\ \hline
    0.007 & $20^3\times 64$ & 0.1842(7)  & 0.696(7) & 292.99(111) & 1107.1(111) \\
    0.010 & $28^3\times 64$ & 0.2238(5)  & 0.726(5) & 355.98(80)  & 1154.8(80)  \\
    0.010 & $20^3\times 64$ & 0.2238(5)  & 0.726(5) & 355.98(80)  & 1154.8(80)  \\
    0.020 & $20^3\times 64$ & 0.3113(4)  & 0.810(5) & 495.15(64)  & 1288.4(80)  \\
    0.030 & $20^3\times 64$ & 0.3752(5)  & 0.878(5) & 596.79(80)  & 1396.5(80)  \\
    \hline\hline
  \end{tabular}
  \caption{Physical pion and nucleon masses.}
  \label{tab:phys-pars}
\end{table}

In previous publications we often computed propagators by chopping
each lattice in two halves and performing propagator calculations
independently on both halves, cf.~Ref.~\cite{Hagler:2007xi} and
references therein. In the present work we adopt a different
technology and compute multiple source/sink pairs on a single gauge
field. We find this approach both more convenient and more powerful,
resulting in a superior statistical quality of our results. By
choosing eight different source/sink pairs on a single gauge field, we
managed to reduce our error bars by a factor of two. We also took
possible sources of correlations into account by performing fits using
the error correlation matrix among all data points on each ensemble,
see e.g.~\cite{Syritsyn:2009np}, and the ``super jackknife''
technique, Refs.~\cite{Blum:2009pr, AliKhan:2001tx}, for combining
data from different ensembles in a single fit.

\section{\label{sec:form-factor-results}Form factor results}
We discuss several results of our calculation of nucleon form
factors. To study the shape of the nucleon at large distances ---
which is a property that can be studied well by lattice calculations
--- we perform an expansion of the form factors at small $Q^2$,
yielding the mean square radii, $\langle r^2_i\rangle$, as the slope,
where $i$ denotes either $1$, $2$, or $A$, corresponding to the form
factor $F_1$, $F_2$, or $G_A$, respectively. Phenomenologically, the
Dirac radius $\langle r_1^2\rangle$ can be determined from fits to the
form factor $F_1(Q^2)$~\cite{Amsler:2008zzb} or from an analysis based
on dispersion theory~\cite{Hohler:1976ax, Mergell:1995bf,
  Belushkin:2006qa}. These two methods currently yield inconsistent
results. For $F_2(Q^2)$ there is a systematic discrepancy between
spin-transfer and Rosenbluth experiments, the source is generally
believed to be two-photon exchange processes, see
Ref.~\cite{Arrington:2007ux} and references therein. Lattice
calculations allow for a study without two-photon contamination and
thus can be very useful in resolving this discrepancy.

We have studied the form factors $F_1(Q^2)$ and $F_2(Q^2)$ using
dipole and tripole fits and also fit our lattice data to the
simultaneous expansion in $Q^2$ and $\mps$ obtained from the
small-scale expansion (SSE), see Refs.~\cite{Bernard:1998gv,
  Hemmert:2002uh} for the explicit form of these expressions. The
advantage of the simultaneous expansion is that we do not make
model-dependent assumptions on the $Q^2$ dependence of the form
factors, at variance with the use of dipole or tripole
phenomenological formulae. The disadvantage is that the validity of
the expansion will only hold for small values of $Q^2$ and we have
only few data points in that region. Thus, relying on the SSE
expansion may increase the uncertainty, both statistical and
systematic (which accounts for the unknown magnitude of higher order
contributions). Since it is not feasible to determine all the
low-energy constants involved in the chiral expressions by fitting to
our lattice results, we fix some of them using their phenomenological
values.

\begin{figure}[htb]
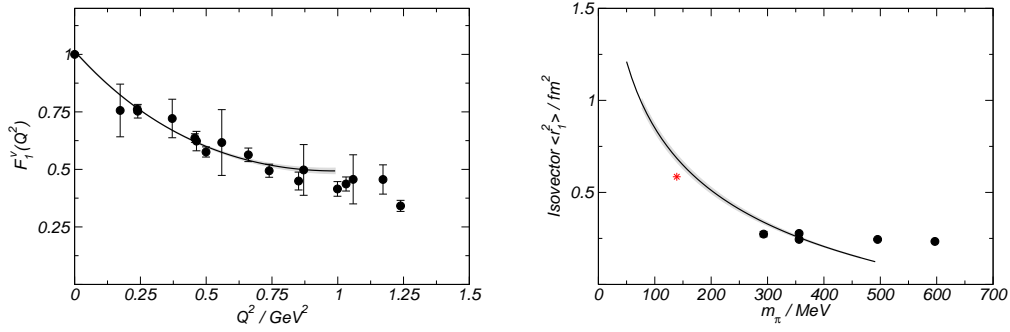

  \centering
  \includegraphics[scale=0.25,clip=true]{F1v_sse.eps}\qquad
  \includegraphics[scale=0.25,clip=true]{r1v_sse.eps}
  \caption{Isovector form factor $F_1(Q^2)$ lattice data with best fit
    SSE at $m_\pi=292.99$ MeV (left panel). The line in the right-hand
    panel shows the resulting Dirac radii, $\langle
    r_1^2\rangle$. Also shown as the data points are the Dirac radii
    obtained from dipole fits to the form factors at each pion mass}
  \label{fig:comb-final}
\end{figure}

We find that applying the cuts $Q^2<0.5$ GeV$^2$ and $\mps<400$ MeV
yields an acceptable fit with $\chi^2/$dof$=1.86$ with two fit
parameters. We could still describe the function well for larger
$Q^2$, but the dependence on $\mps$ is worse. Despite this
observation, we believe that the apparent agreement between the
lattice data and the SSE form for $Q^2>0.5$ GeV$^2$ is merely
accidental since we have no reason to believe that the SSE at the
order given is valid for that range of
$Q^2$. Figure~\ref{fig:comb-final} summarizes our results. The left
panel shows the isovector form factor $F_1(Q^2)$ lattice data with the
best fit SSE at $\mps=292.99$ MeV. The right panel shows the resulting
chiral extrapolation of the Dirac radii as a function of the pion
mass, $\mps$. For illustration purposes, we have also included the
Dirac radii obtained from dipole fits in the graph. However, these
data points have no influence on the curve presented and simply
compare the two fit methodologies. The red star shows the empirical
value taken from Ref.~\cite{Amsler:2008zzb}.

\begin{figure}[htb]
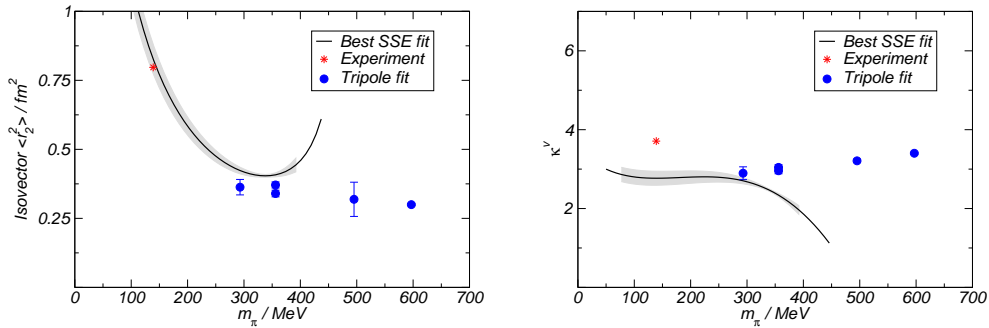

  \centering
  \includegraphics[scale=0.25,clip=true]{r2v_sse.eps}\qquad
  \includegraphics[scale=0.25,clip=true]{kappav_sse.eps}
  \caption{Isovector Pauli radius $\langle r_2^2\rangle$ (left panel)
    and isovector anomalous magnetic moment $\kappa_v$ (right panel)
    as a function of the pion mass.}
  \label{fig:f2v-summary}
\end{figure}

The corresponding fit for the isovector $F_2(Q^2)$ had a quality of
$\chi^2/$dof$=1.31$ with four fit parameters when applying the cuts
$Q^2<0.3$ GeV$^2$ and $\mps<400$ MeV. However, we noticed that the
$Q^2$ dependence of the SSE expression was in worse agreement than in
the case of $F_1(Q^2)$, while the $\mps$ dependence was better. The
resulting Pauli radii and anomalous magnetic moments are shown in
Fig.~\ref{fig:f2v-summary}. The left panel shows the Pauli radius
$\langle r_2^2\rangle$, while the right panel shows the isovector
anomalous magnetic moment, $\kappa_v$. Again, data points from tripole
fits are included in the plot, but have no influence on the fit.
We find that $\langle r_2^2\rangle$ is described well by the fit and
the fit even gets close to the experimental point. The magnetic moment
$\kappa_v$ is lower than the experimental value. We find these fits
encouraging, but believe that the excellent agreement with experiment
is accidental for $\langle r_2^2\rangle$. Data with smaller pion mass
are needed to verify this finding since $\kappa_v$ does not yet agree,
although it originates from the same data set.

\begin{figure}[htb]
  \centering
  \includegraphics[scale=0.25,clip=true]{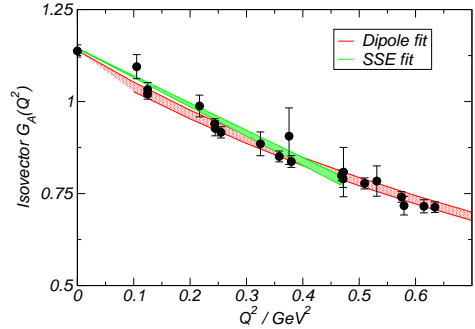}
  \caption{Comparison of dipole and chiral fit to $G_A(Q^2)$ for the
    $28^3$ lattice at $m_\pi=355.98$ MeV.}
  \label{fig:ga-sse-dip-comp}
\end{figure}

Similar to our fit strategy for the vector form factors, we also adopt
a simultaneous fit to the $Q^2$ and $\mps$-dependence of the axial
form factor, $G_A(Q^2)$. Figure~\ref{fig:ga-sse-dip-comp} shows the
result of the chiral fit together with the dipole fit and the data set
for the $28^3$ lattice at $\mps=355.98$ MeV with a fitting range of
$Q^2<0.4$ GeV$^2$ for the SSE expansion and all $Q^2$ values for the
dipole fit. The SSE fit gives a $\chi^2/$dof$=1.73$. The resulting
axial radius, however, is $\langle r_A^2\rangle = 0.1560(60)$ fm$^2$
for the cuts $Q^2<0.4$ GeV$^2$ and $\mps<400$ MeV, substantially lower
than the experimental value in Ref.~\cite{Bernard:2001rs}. Future
lattice calculations at smaller pion masses will be crucial to resolve
this issue.

The induced pseudoscalar form factor, $G_P(Q^2)$, is not described by
a dipole type fit formula. Instead, it is commonly fit using a
pion-pole expression, giving excellent agreement with the data, see
Ref.~\cite{Bernard:2001rs}. We performed two kinds of fits: First, we
repeated the analysis done previously with the other form factors,
i.e.~performing a combined fit in $Q^2$ and $\mps$. Second, we took
the pion-pole form as a function of $Q^2$ and fit it using a single
ensemble with fixed $\mps$, treating the pion mass as a free
parameter.

\begin{figure}[htb]
  \centering
  \includegraphics[scale=0.25,clip=true]{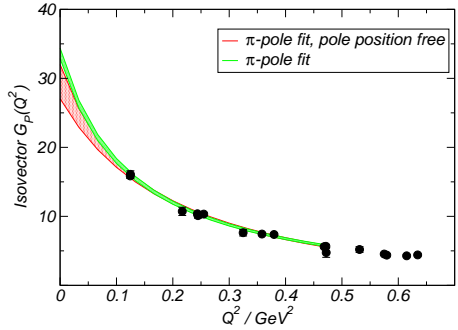}
  \caption{Comparison of pion-pole fits to isovector $G_P(Q^2)$ on the
    $28^3$ lattice with fixed pion pole and with the pion pole as a
    free parameter.}
  \label{fig:gp-pole-comp}
\end{figure}

In the first case, we again find that kinematic cuts of $Q^2<0.5$
GeV$^2$ and $\mps<400$ MeV yield reasonable results. In the second
case, we find the location of the pion pole, $\mps=417(43)$ MeV, with
an uncertainty of $10\%$ within the actual value of $\mps=355.98$ MeV
on the $28^3$ lattice. The resulting $\chi^2/$dof$=0.94$ indicates an
excellent fit to the data. Figure~\ref{fig:gp-pole-comp} shows a
comparison of the two fits. It is evident that both fits manage to
describe the data well, but the uncertainty of the curve with $\mps$
as a free parameter is larger.
We conclude that the isovector $G_P(Q^2)$ is described well by the
pion-pole form.

\section{\label{sec:summary-outlook}Summary and Outlook}
We have computed the isovector form factors of the vector and the
axial current for the nucleon within our framework of mixed action
calculations. We have applied new analysis techniques that result in
substantially reduced error bars at minimal additional computational
cost. Furthermore, due to the agreement between the lattice results
reported in this proceeding and the corresponding one reported
in~\cite{Meifeng:2009lat} obtained using full DWF, we are confident
that the hybrid calculations employed do not suffer from systematic
effects.

We find that a combination of chiral fits and lattice data is possible
with the current generation of lattice calculations. This way, we
obtain qualitative agreement with many features we expect to hold when
approaching the chiral limit. We expect that the upcoming generation
of lattice calculations will provide conclusive quantitative results
from first principles that will be in agreement with experiment
without resorting to assumptions on functional behavior outside of
what can be predicted by chiral perturbation theory. We are able to
provide fits to the vector form factors, $F_1(Q^2)$ and
$F_2(Q^2)$. While we have no explanation for the discrepancy of the
axial radius, $\langle r_A^2\rangle$, we find that induced
pseudoscalar form factor is described well by the pion-pole form.

\acknowledgments This work was supported in part by U.S.~DOE Contract
No.~DE-AC05-06OR23177, by the DOE Office of Nuclear Physics under
grants DE-FG02-94ER40818, DE-FG02-04ER41302, DE-FG02-96ER40965,
DE-FG02-05ER25681 and DE-AC02-06CH11357 and the EU (I3HP) under
contract No.~RII3-CT-2004-506078. Ph.H.~and B.M.~acknowledge support
by the Emmy-Noether program and the cluster of excellence ``Origin and
Structure of the Universe'' of the DFG and W.S.~acknowledges support
by the National Science Council of Taiwan under the grant numbers
NSC96-2112-M002-020-MY3 and NSC96-2811-M002-026 and wishes to thank
the Institute of Physics at Academia Sinica for their kind hospitality
and support as well as Jiunn-Wei Chen at National Taiwan University
and Hai-Yang Cheng and Hsiang-Nan Li at Academia Sinica for their
hospitality and for valuable physics discussions and
suggestions. K.O.~acknowledges support from the Jeffress Memorial
Trust grant J-813 and Ph.H., M.P.~and W.S.~acknowledge support by the
A.v.~Humboldt-foundation through the Feodor-Lynen program. This
research used resources under the INCITE and ESP programs of the
Argonne Leadership Computing Facility at Argonne National Laboratory,
resources provided by the William and Mary Cyclades Cluster, and
resources provided by the New Mexico Computing Applications Center
(NMCAC) on Encanto. These calculations were performed using the Chroma
software suite~\cite{Edwards:2004sx}. We are indebted to members of
the MILC Collaboration for providing the dynamical quark
configurations that made our full QCD calculations possible.


\end{document}